\documentclass[%
aip,
amsmath,amssymb,
reprint,nofootinbib%
]{revtex4-2}
\usepackage{nomencl}
\makenomenclature

\usepackage{graphicx,array,booktabs,rotating}
\usepackage{dcolumn}
\usepackage{bm}
\usepackage{mathptmx}
\usepackage{multirow}
\usepackage{afterpage,float,color,xcolor}
\usepackage{hyperref}
\usepackage{etoolbox} 
\usepackage{tabularx,tabulary}
\usepackage{tikz}
\newcommand\encircle[1]{%
  \tikz[baseline=(X.base)] 
    \node (X) [draw, shape=circle, inner sep=0] {\strut #1};}
\usepackage[capitalize]{cleveref}
\usepackage{times}
\hypersetup{
	colorlinks,
	linkcolor={blue!100!black},
	citecolor={blue!100!black},
	urlcolor={blue!100!black}
}
\newcommand{\PRLsep}{\noindent\makebox[\linewidth]{\resizebox{0.3333\linewidth}{1pt}{$\bullet$}}\bigskip}
\makeatletter
\newcommand\footnoteref[1]{\protected@xdef\@thefnmark{\ref{#1}}\@footnotemark}
\makeatother

\newcommand{\etal}{\textit{et al.}}
\usepackage[page]{totalcount}
\usepackage{etoolbox,fancyhdr,xcolor}%
\begin{document}
\preprint{AIP/123-QED}
\title[\textbf{Phys. Fluids} (2023) $|$  Manuscript under preparation]{On the flow unsteadiness and operational characteristics of a novel supersonic fluidic oscillator}

\author{Spandan Maikap}%
\email{maikap.1@iitj.ac.in}
\affiliation{ 
Department of Mechanical Engineering, \\Indian Institute of Technology Jodhpur, Jodhpur-342030, Rajasthan, India}

\author{S. K. Karthick}%
\email{skkarthick@ymail.com}
\affiliation{ 
Department of Mechanical Engineering, \\Indian Institute of Technology Madras, Chennai-600036, Tamilnadu, India}

\author{R. Arun Kumar}%
\email{arunkr@iitj.ac.in (corresponding author)}
\affiliation{ 
Department of Mechanical Engineering, \\Indian Institute of Technology Jodhpur, Jodhpur-342030, Rajasthan, India}

\date{\today}

\begin{abstract} 
A novel supersonic jet oscillating method is investigated both experimentally and numerically. A rectangular primary supersonic jet is issued into a confined chamber with sudden enlargement. Secondary control jets are issued from the top, and bottom  backwards-facing step regions formed due to sudden enlargement. The secondary jet also expands in the confined chamber shrouding the primary jet from the top and bottom sides.
The primary jet is oscillated in the transverse direction by blowing the secondary jets in the streamwise direction in a pulsating manner with a phase shift. The out-of-phase secondary jet blowing causes the primary jet to periodically adhere to the upper and lower part of the confined chamber, causing flapping of the primary jet and acting as a supersonic fluidic oscillator. The supersonic jet oscillation characteristics are experimentally investigated using shadowgraph type flow visualization technique and steady and unsteady pressure measurements. Quantitative analysis of the shadowgraph images using the construction of $y-t$ and $y-f$ plots reveals the presence of periodic jet oscillation with a discrete dominant frequency similar to the secondary jet excitation frequency. The existence of linearity between the excitation frequency and the flapping jet frequency on the low-frequency ($0.66-6.6$ Hz) side is first proven experimentally. Later, the high-frequency ($16.67-1666.66$ Hz) operation extent of the supersonic fluidic oscillator is further demonstrated using unsteady computational studies owing to the existing experimental facility's limitations. It is found from the computational studies that there exists a limiting oscillation frequency for the present fluidic oscillator (nearly 5000 Hz with the particular geometric size and the injection momentum considered in the present study). A reduced-order analytical framework based on Newton's second law has also been proposed to investigate the limiting oscillation frequency. It is found that the limiting frequency predicted from the proposed analytical model shows fairly good agreement with the computationally predicted results.

\end{abstract}

\keywords{gas dynamics, unsteady flow, flow control, supersonic fluidic oscillator} 
\maketitle

\section{Introduction}\label{sec:intro}
A fluidic oscillator\cite{Wen2020,Veerasamy2022,Ayeni2022} is a device that can oscillate a fluid flow in a controlled manner. Such devices find immense applications in manipulating the flowfield for various engineering applications, such as controlling the aerodynamic characteristics of airfoils \cite{Xia2021,Kim2020,DeSalvo2010,Shmilovich2017}, controlling the jet engine exhaust \cite{Kibens1999}, acoustic tone reduction \cite{Raman2004}, heat transfer enhancement \cite{Wu2019}, and so on. The overwhelming requirement for fluidic oscillators in various applications prompted numerous research in this field. Several promising active and passive flow control strategies for creating an oscillating jet have been developed. Many of these fluidic control strategies show excellent control abilities at low speeds\cite{Arote2019}. However, effective flow control at high speed still poses many challenges due to the overwhelming distribution of flow momentum in one particular direction. Developing fluidic oscillators for high-speed applications faces various complexities and requires further research to develop more feasible and economical fluidic oscillators.

\begin{figure*}
	\includegraphics[width=0.8\textwidth]{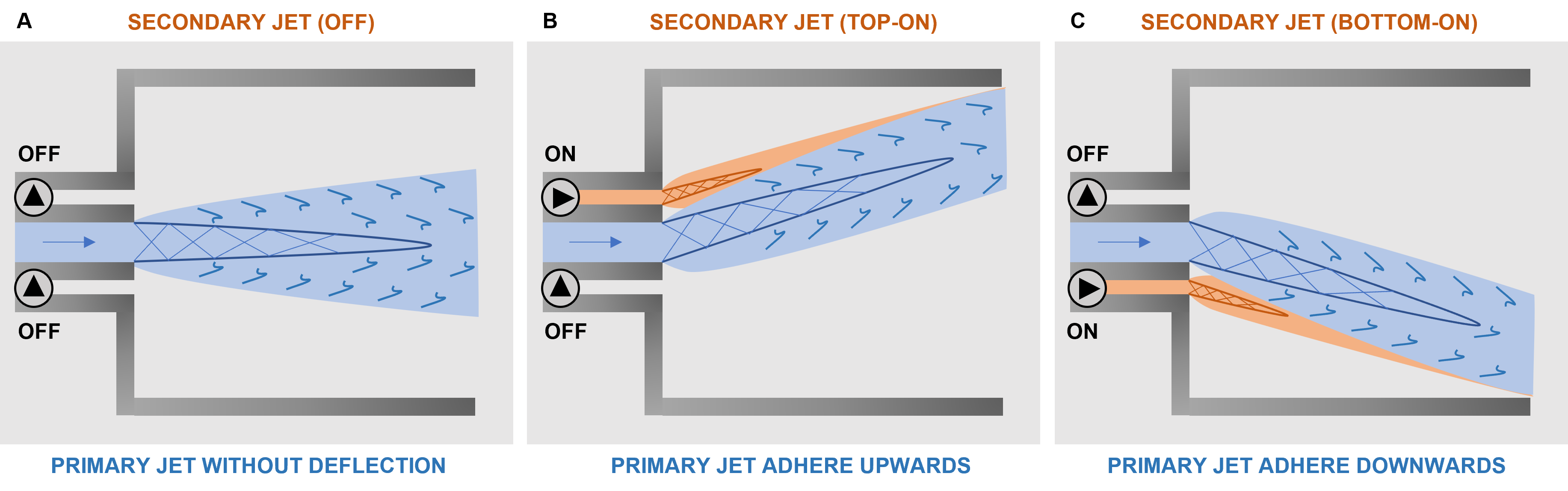}
	\caption{\label{fig:mechanism_schematic} Schematic explaining the idea behind the establishment of a supersonic fluidic oscillator. (a) When there is no secondary jet, the primary jet goes unaffected; When the secondary jet in the top (b) or bottom (c) is turned on, the primary jet deflects and adheres to the upper or lower wall, respectively.}
\end{figure*}

One of the pioneering works in fluid oscillators was reported by  Viets\cite{Viets1975}  using a flip-flop jet oscillator. A flip-flop jet oscillator uses a feedback tube that connects the jet's top and bottom portions as it expands from the nozzle. Due to the bistable nature of the jet, it sticks to either the top or bottom wall, thereby creating a pressure differential in the feedback loop. A pressure wave is established in the feedback loop, which pushes the jet to the other wall. This process cyclically repeats itself, creating a self-sustained oscillating jet. The flip-flop jet's adaptability in oscillating a supersonic jet was investigated by Raman \etal \cite{Raman1993}. They found that the flip-flop jet nozzle shows oscillatory behaviour up to a critical Mach number for a fixed geometry. Changing the geometrical parameters can resolve the requirement for increasing the Mach number. Due to the advantage of self-sustained oscillatory jets, flip-flop nozzles found application in subsonic mixing enhancement\cite{Parekh1996} and heat transfer enhancement in impinging jets\cite{Camci2002}. Another study on fluidic oscillators in which a supersonic jet was injected into a dome-shaped\cite{Tomac2018} chamber was investigated by Raghu\cite{Raghu1999}. Due to the instability of the vortex structures inside the chamber, the jet flutters periodically, creating an oscillating jet. Other fluidic oscillators, such as jet-wedge-resonator oscillators and edge-tone oscillators\cite{Campagnuolo1969}, work on the principle of acoustic tones produced by an airstream moving on a wedge/edge (Aeolian tone). The jet oscillation frequency in such fluidic oscillators depends on the distance between the nozzle and wedge, limiting their ability to supply on-demand oscillation frequencies.

Gokoglu \etal \cite{Gokoglu2011} investigated a feedback-loop-based fluidic oscillator where a feedback loop is established between the upstream and downstream portions of the jet. The two-dimensional (planar) nature of the fluidic oscillator's nozzle makes it more feasible and applicable in fluidic systems than the flip-flop oscillator\cite{Viets1975}, which has a 3D geometry. These fluidic oscillators could provide a 1 to 10 kHz jet oscillation frequency for a meagre mass flow rate\cite{Raghu2013} of 1 g/s. It was observed that for a fluidic oscillator, the oscillation frequency could only be increased with an increase in Mach number up to a critical limit, after which the oscillator performance decreased significantly\cite{Gokoglu2011}. Vatsa \etal \cite{Vatsa2012} further investigated the supersupersonic fluidic oscillator by modifying the fluidic oscillator geometry to enhance the supersonic oscillating jet. They found that fluidic diverters with smoothened corners have a better spread rate than those with sharp corners. The internal flow features of such fluidic oscillators were investigated by Bobusch \etal \cite{Bobusch2013}. They found that the growing separation bubble between the main jet and the nozzle wall is the underlying reason for the oscillation mechanism. 

Sang \etal \cite{Sang2020} investigated the effect of geometrical parameters on a fluidic oscillator with supersonic flows. They found that the nozzle and throat width significantly affect the initiation time and the oscillation period. Xu \etal \cite{Xu2013} investigated the performance of a widely used feedback-based fluidic oscillator\cite{Tesa2006,Yang2007} and found a characteristic relation between the flow Reynolds number ($Re$) and the oscillation frequency. Tesař \etal \cite{Tesa2013} investigated the fluidic oscillator by substituting the feedback loop with a quarter-wave Helmholtz resonator. They found that the newly developed fluidic oscillator had an advantage over the conventional oscillator by having a higher frequency. Moreover, the frequency did not depend on the jet flow rate. Tesař \etal \cite{Tesa2015} also investigated another high-frequency fluidic oscillator based on a stationary rotating vortex between the two walls. Such an oscillator is advantageous as it does not contain moving parts or feedback channels. It achieves an oscillating flow with a frequency of up to 8.2 kHz for a flow $Re$ between 2300 to 6000. More detailed descriptions and classifications of different fluidic actuation devices are also available in the works of Cattafesta and Sheplak\cite{Cattafesta2011}, and Wang \etal \cite{Wang2012}.

Due to the superiority of the fluidic diverters actuators over their counterparts, they found several applications in the field of aerodynamics \cite{Phillips2010,Ostermann2019,Ghanami2019,Feikema2008,Maikap2022}. Phillips \etal \cite{Phillips2010} investigated the use of fluidic diverter actuators for flow separation delay in airfoil wings. Ostermann \etal \cite{Ostermann2019} showed that a fluidic diverter actuator could be applied in crossflow mixing enhancement. Dominant flow patterns in a crossflow, such as the counter-rotating vortex pairs, showed unsteadiness, increasing entrainment of the surrounding fluid and resulting in better mixing. Raman \etal \cite{Raman2004} found that by using the fluidic diverters, cavity resonance tone can be suppressed by as much as 10 dB, while injection of a simple fluid mass into the cavity flow suppressed the tones by only 1 dB. 

Ghanami and Farhadi \cite{Ghanami2019} investigated the feedback loop-based fluidic diverter for heat transfer enhancement on a heated pipe. Their research concluded that installing fluidic oscillators on the pipe wall significantly increased heat transfer. Feikema \etal \cite{Feikema2008} studied a fluidic oscillator that injects two control jets transversely into the main jet. They found that the fluidic oscillator works efficiently up to a critical frequency of 250 Hz, after which the jet oscillation ceases. 

\begin{figure*}
	\includegraphics[width=0.8\textwidth]{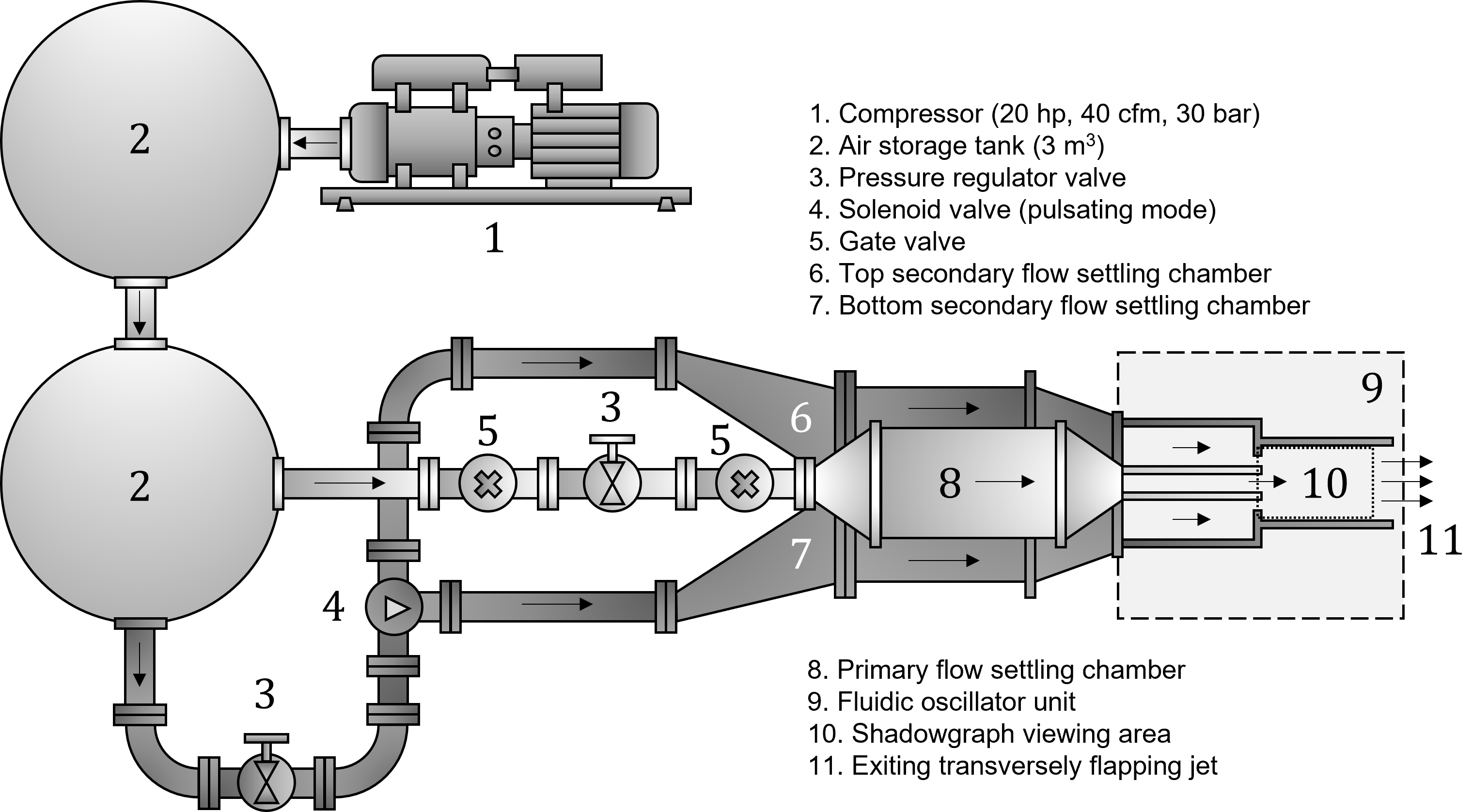}
	\caption{\label{fig:schematic} A typical schematic of the blow-down facility used to develop and study the supersonic fluidic oscillator for a range of low-frequency actuation along with the associated blow-down and pipe assembly units. Key specifications of the blow-down units are given in the schematic's annotations. The schematic is not drawn to the scale and does not directly depict the actual installation. Flow exits from left to right.}
\end{figure*}

Maikap \etal \cite{Maikap2022} did a preliminary numerical investigation on a fluidic oscillator where a sonic jet (primary) has oscillated using two pulsating sonic jets (secondary) in the streamwise direction due to the Coanda effect. The underlying flow events are schematically redrawn in Figure \ref{fig:mechanism_schematic}. They demonstrated that the mass flow rate of the two jets and the actuation frequency significantly affect the sonic jet's oscillation characteristics. 

The brief literature review shows that many successful fluid oscillators have been developed for flow control at subsonic speed. However, oscillating a supersonic jet using conventional fluid oscillators still faces many challenges, and only limited success has been achieved. For instance, on-demand frequency control of the fluidic oscillator is a significant challenge. In flip-flop jet oscillators, the tube length must be changed to alter the oscillation frequency, which is impractical while the system is undergoing operation\cite{Viets1975}. Fluidic diverter-based actuators have a critical Mach number up to which an efficient supersonic oscillation is obtained. Moreover, the oscillation frequency also depends upon the choice of geometry\cite{Gokoglu2011}. The control strategy of the supersonic fluidic oscillator by Maikap \etal \cite{Maikap2022} carries only preliminary computational results for a limited actuation frequency.

In the present study, the authors aim to experimentally investigate a field-employable supersonic fluidic oscillator as a proof-of-concept device. Moreover, the on-demand frequency requirements, limiting operating conditions and the driving flow physics of such a device are targeted to be studied in detail using computational methods. Such a broad knowledge of the proposed supersonic fluidic oscillators may find applications in jet thrust vectoring, heat transfer enhancement by impinging jets and mixing enhancement in cross-flows.  

The rest of the manuscript is organized in the following manner: In Sec. \ref{sec:exp_meth}, a brief description of the adopted experimental methodology is provided. In Sec. \ref{sec:num_meth}, the computational methodology used for the present study is described. In Sec. \ref{sec:res_disc}, results and discussions from the experimental and computational studies are given under different subsections. The proof-of-concept of the device is experimentally proven between Sec. \ref{ssec:gross_flow} and Sec. \ref{ssec:flow_phys}. Using computational methods, the operation characteristics at high-frequency actuation are elaborated in Sec. \ref{ssec:num_study}. The limiting operation of the fluidic oscillator is discussed using an approximated model at Sec. \ref{ssec:lim_op}.  At the end in Sec. \ref{sec:conclusions}, some of the vital findings from the present study are listed. 

\section{Experimental Methodology}\label{sec:exp_meth}

All the experiments are performed in the open-jet facility available at the Shock Waves and High-Speed Flow Lab in IIT Jodhpur-India, as shown in the schematic of Figure \ref{fig:schematic}. The facility comprises primary and secondary jet feed lines drawn from the central air storage units. The primary feed line is directly connected to the primary jet flow settling chamber, where the total pressure ($p_{0p}$) is measured at the end. The secondary feed line is further split and connected to the top and bottom portions of the secondary jet flow's settling chamber, where the total pressure ($p_{0s}$) is measured at the end. The primary and the secondary flow are provided with an equivalent stagnation pressure of $p_0=p_{0p}=p_{0s}=3$ bar at the operation time, using pressure regulating valves placed at the primary and secondary feed lines. The secondary jet flow from the top and the bottom settling chamber is actuated periodically using a fast-acting solenoid valve (UNI-D, UV-10) during the supersonic fluidic oscillator's operation. The solenoid valve opening and closing are controlled using a NI\textsuperscript{\tiny\textregistered}-9478A sinking output module. A LabVIEW\textsuperscript{\tiny\textregistered} code has been developed such that the NI\textsuperscript{\tiny\textregistered} 9478 generates an alternating pulsed signal supplied to the solenoid valve for the necessary actuation. 

\begin{figure*}
	\includegraphics[width=0.9\textwidth]{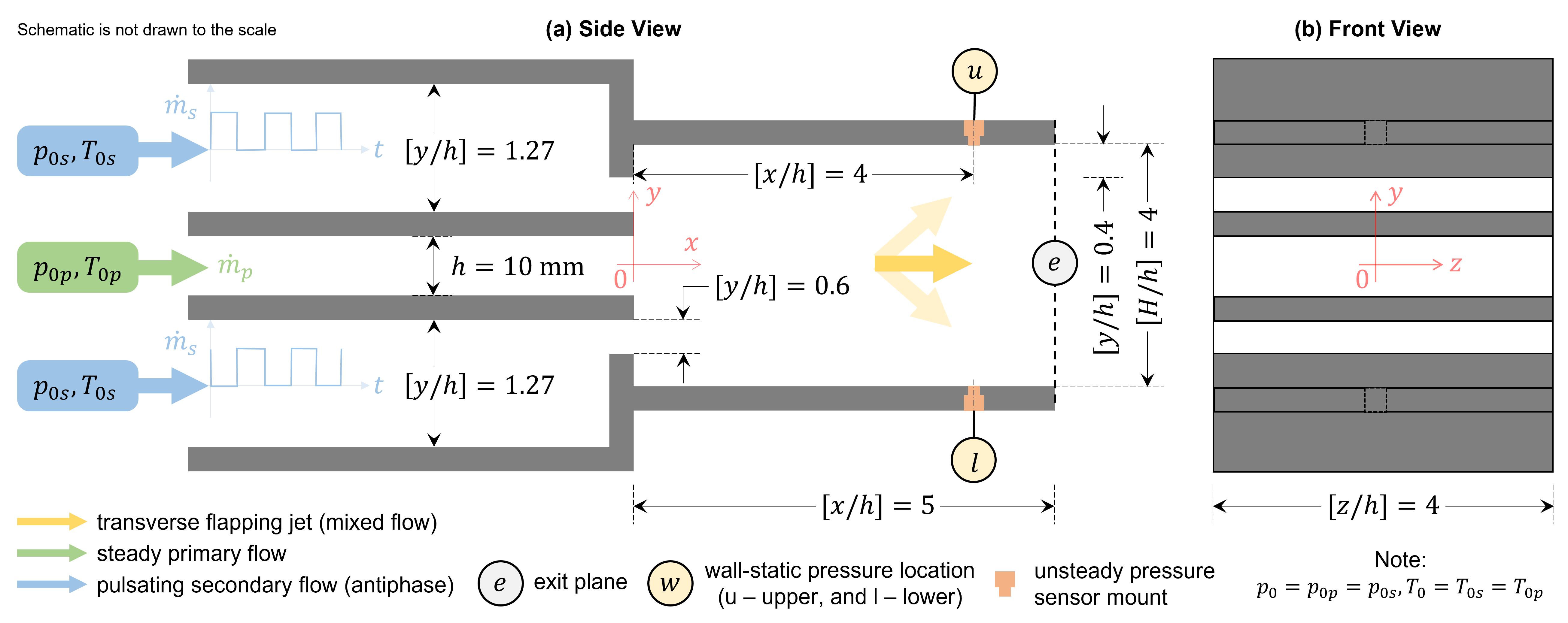}
	\caption{\label{fig:model_details} A typical schematic showing the generic operation of the sonic jet oscillator used in the present experiments, along with the defining geometric scales, flow conditions, respective axis system, and measurement stations. The schematic is not drawn to the scale. The secondary flow is actuated in step pulses, and the primary flow is steady. The exiting mixed flow is flapping transversely as a response to the actuation in the $xy$ plane.}
\end{figure*}

The primary core flow and the secondary flows on the top and bottom enter a confined chamber through constant area nozzle openings (see Figure \ref{fig:schematic}). The nozzle opening area in the $yz$ plane for the primary and the secondary flow is $10 \times 40$ mm$^2$ and $6 \times 40$ mm$^2$, respectively. Moreover, two backwards-facing steps of 15 mm height and 40 mm width are provided on the top and bottom sides of the secondary flow nozzle. The confined chamber into which those nozzles are attached has a cross-sectional area of $40 \times 40$ mm$^2$ in the $yz$ plane and has a length of $50$ mm. Flows expand through those nozzles in the confined chamber, forming under-expanded jets\cite{ArunKumar2018,Raju2022} provided the nozzle pressure ratio ($\zeta=p_0/p_a$) is sufficient to choke the exiting flow from the respective nozzles. Figure \ref{fig:model_details} gives more details about the flow geometry, origin, axes and measurement stations.

The jet oscillation characteristics using the present control strategy are further investigated qualitatively through a high-speed flow visualization technique and quantitatively by pressure measurements. Visualization is done using a `z-type' shadowgraph arrangement\cite{Settles2001}. The light source for the visualization is achieved from a 50 W halogen bulb. The light from the bulb is further focused on a rectangular slit using a condensing lens to generate an extended light source. The extended light source is kept at the focus of the first field mirror (concave mirror of 200 mm diameter and 1000 mm focal length) to produce a collimated light beam that illuminates the test section. Time-varying shadowgraph images of the test section are imaged using a high-speed camera (Chronos\textsuperscript{\tiny\textregistered} 2.1 HD). The images are recorded at 1000 frames per second (fps), with a light exposure time of 10 $\mu$s. The images are obtained at the largest frame size of $1920 \times 1080$ pixels and at a pixel resolution of 0.0558 mm/pixel.

Pressure at any point along the flow except at the wall of the confined chamber is determined by placing a pitot probe at respective locations. The pitot and wall-static pressure are measured using a piezo-resistive pressure sensor of Keller\textsuperscript{\tiny\textregistered} make (M5HB type with accuracy: $\pm 0.1 \%$ FS). The pressure sensors are connected to the pressure ports via polyurethane pipes to record the pressure data at a sampling rate of 50 kHz. The open-jet facility's settling chamber pressures are measured using a Keller\textsuperscript{\tiny\textregistered} based pressure sensor PAA-21Y (accuracy: $\pm 0.25 \%$ FS) with a limiting frequency of 2 kHz. When deflected to any one wall side, the supersonic jet is expected to produce a pressure peak due to jet impingement. The pressure measurement at the upper and lower wall near the jet exit can reveal the jet's oscillation characteristics. Two unsteady pressure ports, in station \encircle{$u$} (upper) and \encircle{$l$} (lower), have been provided (see Figure \ref{fig:model_details}) to extract the pressure fluctuations.

\begin{table}
	\caption{\label{tab:flow_cond} Tabulation of the flow conditions realized in the experimental and computational campaign in the study of the supersonic fluidic oscillator.}
	\begin{ruledtabular}
		\begin{tabular}{lc}
			\textbf{Parameters \footnote{The fully expanded jet properties, the total flow conditions, and the ambient conditions are marked with subscripts $j, 0,$ and $a$, respectively. The quantities are calculated based on $\zeta$.}} & \textbf{Values} \\ \midrule
            Ideal nozzle pressure ratio\footnote{In experiments, the back pressure will be different due to the confinement's presence around the primary jet.} ($\zeta = p_0/p_a$) & 3 \\
            Total pressure\footnote{\label{fn_1} The total pressure and temperature in the primary and secondary flow are the same ($p_0=p_{0p}=p_{0s}$ and $T_0=T_{0p}=T_{0s}$).} ($p_0 \times 10^5$, Pa) & 3\\
            Total temperature\footnoteref{fn_1} ($T_0$, K) & 300\\
            Fully expanded jet velocity ($u_j$, m/s) & 403 \\
            Fully expanded jet height ($h_j\times 10^{-3}$, m) & 10.9 \\
            Fully expanded jet Mach number ($M_j$) & 1.358\\
            Fully expanded jet Reynolds number ($Re_{h_j} \times 10^5$) & 4.83\\
            Blowing ratio\footnote{Calculated based on the measured $p_0$ and exit area of the nozzle as specified in Figure \ref{fig:model_details} using isentropic relations.} ($\eta=\dot{m_s}/\dot{m_p}$) & 0.6\\
		\end{tabular}
	\end{ruledtabular}
\end{table}

\section{Numerical Methodology} \label{sec:num_meth}

\begin{figure*}
	\includegraphics[width=0.9\textwidth]{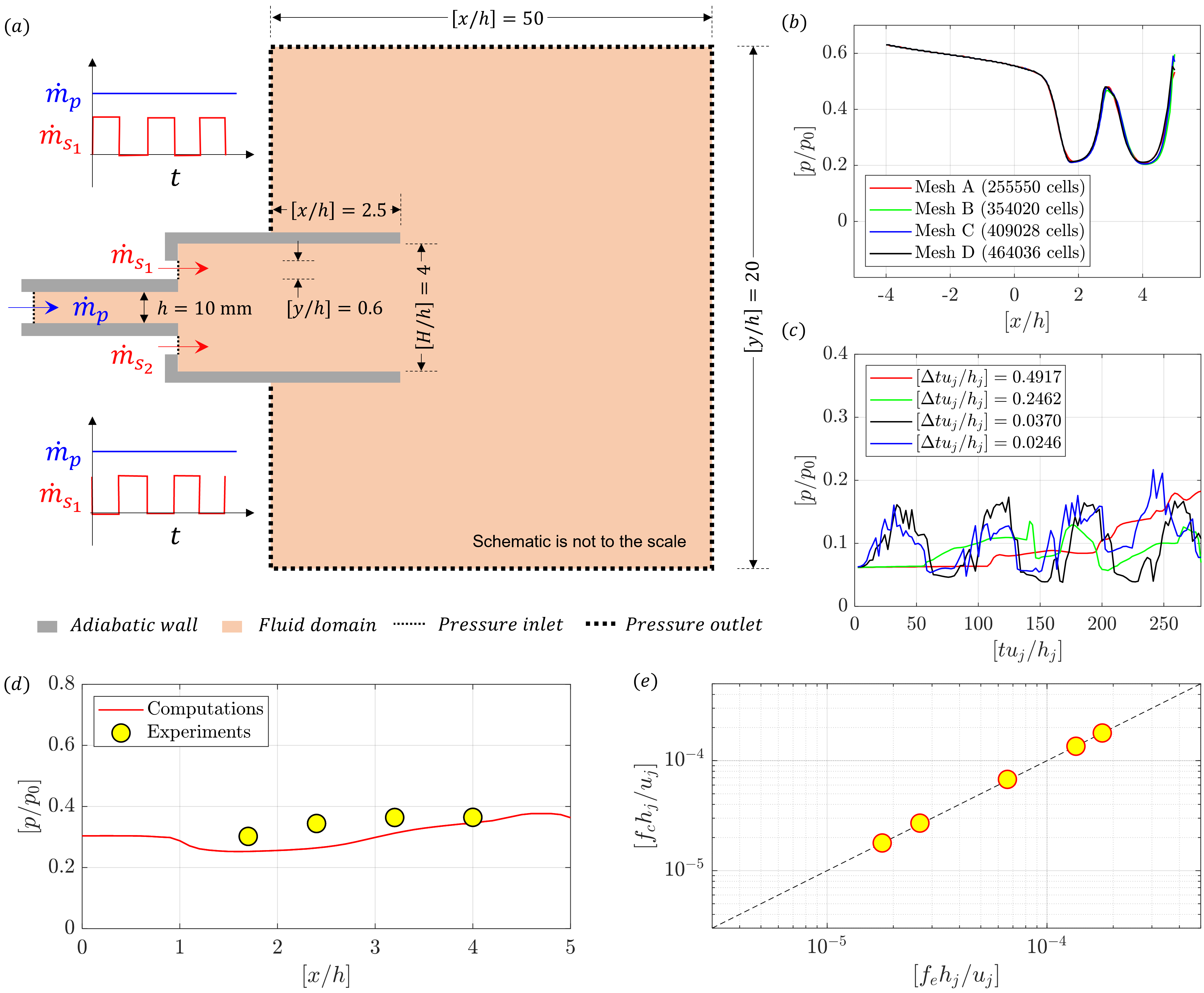}
	\caption{\label{fig:cfd_merge}(a) A typical CFD domain considered in the present computational campaign of the supersonic fluidic oscillator, along with the boundary conditions and the geometrical extents. The origin is at the centre as mentioned in the schematic of Figure \ref{fig:schematic}; (b) Graph showing the mesh independence studies through the plot of centerline ($y/h=0$) non-dimensionalized static pressure variation for four different grid densities with only primary flow turned-on; (c) Plot showing the time independence studies by monitoring the non-dimensionalized static pressure variation over time at $[x/h,y/h]=[2,0]$ for four different non-dimensional time steps (case: C-X); (d) Steady-state wall-static pressure variation (top-wall) between the experiments and computations as part of the solver validation studies for the C-0 case (steady primary and secondary flow); (e) Matching of the dominant transverse jet flapping frequency between the experiments ($f_e$) and computations ($f_c$) for cases between C-I and C-V (see Table \ref{tab:exp_comp_cases}) after measuring $p_{0,e}$ at $[x/h,y/h]=[5,0]$ as part of the solver validation studies.}
\end{figure*}

\subsection{Computational Domain and Numerical Schemes} \label{ssec:comp_dom}
In order to simulate the flowfield, CFD simulations are carried out by solving the two-dimensional compressible unsteady Reynold's Averaged Navier Stokes (URANS) equations. The working medium in the present study is considered air, whose state equation is solved using the ideal gas equation. The viscosity variation due to the temperature change is calculated using Sutherland's approach. The flow turbulence is modelled using the $\kappa-\omega$ SST turbulence model, where separate transport equations are solved for turbulence kinetic energy $`\kappa'$ and the specific dissipation rate $`\omega'$. The governing equations of mass, momentum, and energy were solved in a coupled manner. The cell centre values are extrapolated to the face centres using a second-order upwind scheme. The convective fluxes across the cells are determined by Roe-Flux difference splitting, and the gradients are calculated using the least square-based method. The governing equations are temporally discretized using a second-order implicit formulation. 

The domain of the numerical model consists of a primary duct suddenly expanding into a larger duct, with two backwards-facing steps on the top and bottom side of the primary duct exit. The primary duct inlet is given a pressure inlet boundary condition. The secondary jets are injected from the top and bottom backwards-facing step region, as shown in \ref{fig:cfd_merge}. The secondary jet opening is further subdivided into 15 smaller stepped openings to mimic the gradual opening of the valves. Such procedure of gradual opening of valves has been used in the previous literature \cite{Arun2013}. Pressure inlet boundary condition is provided for both primary and secondary jets with a stagnation pressure and temperature of $p_0$ and $T_0$ as mentioned in Table \ref{tab:flow_cond}. The top and bottom secondary jets are opened and closed cyclically, creating pulsed jets on the top and bottom of the primary jet. The sudden expansion duct exits to a larger domain with a pressure outlet boundary condition having a static pressure of $p=101.325$ kPa, which is ambient.

\subsection{Mesh and time-step independent study}
A rigorous campaign of mesh and time-step independence study is undertaken to converge on the grid size and time step for the final simulations. Firstly, a mesh independence study is carried out in a unique simulation case, where only the primary flow is turned on. The centerline static pressure distribution is plotted for four mesh sizes: 1. Mesh A (255550 cells), 2. Mesh B (354020 cells), 3. Mesh C (409028 cells), and 4. Mesh D (464036 cells). Structured quadrilateral grids are used to discretize the computational domain. The boundary conditions and the numerical schemes used are the same as in the previous section (see Sec. \ref{ssec:comp_dom}). Figure \ref{fig:cfd_merge}b shows the normalized static pressure variation ($p/p_0$) for all four different meshes, and there is no significant variation between Mesh C and D. Hence, Mesh C is considered for the rest of the simulations owing to less memory. 

Secondly, a time-step independence study is performed by simulating the oscillation of the primary jet with the secondary jet actuation. A secondary jet actuation frequency of $f_a=250$ Hz corresponding to the case C-X is considered. The case is run with four different non-dimensional time steps: 1. $[\Delta t u_j/h_j] = 0.4917$, 2. $[\Delta t u_j/h_j] = 0.2462$, 3. $[\Delta t u_j/h_j] = 0.0370$, and 4. $[\Delta t u_j/h_j] = 0.0246$. The time history of the non-dimensional static pressure fluctuation near the jet exit ($[x/h,y/h]=[2,0]$) is monitored for all the cases as shown in Figure \ref{fig:cfd_merge}c. The variations between the cases of $[\Delta t u_j/h_j] = 0.0370$ and 0.0246 are minimal. Hence, a non-dimensional time-step of $[\Delta t u_j/h_j] = 0.0370$ is selected for the rest of the simulations owing to saving the computational cost.

\subsection{Solver validation}
Although a commercial solver is used, a validation study is further done to evaluate its fidelity to the present problem statement. Firstly, one of the flow variables changes across different spatial points are validated with the experiments. The experiments ' wall-static pressure (top wall) is compared with the numerical counterpart for case: C-0. Figure \ref{fig:cfd_merge}d depicts the closely varying trend. Secondly, the primary jet oscillation frequency with various secondary jet actuation frequencies is computationally predicted and compared with the corresponding experimental observations. The details regarding the methodology for determining the primary jet oscillation frequency are available in the upcoming section (see Sec. \ref{ssec:num_study}). The dominant primary jet frequency predicted from the computations closely matches the experiments, as shown in Figure \ref{fig:cfd_merge}e. The validation study thus clearly shows that the adopted commercial solver can predict the proposed fluid oscillator's steady and unsteady flowfield sufficiently.

\section{Results and Discussions} \label{sec:res_disc}

The experimental results produced by the cyclic actuation of a secondary jet on the top and bottom side of the primary jet and the resulting fluidic actuation are discussed in the present section. Moreover, the secondary jet's actuation frequency effect on the primary jet oscillation is also explored. The primary and secondary jet is supplied with a total pressure of $p_{0p}$ and $p_{0s}$, which chokes the jet and causes it to under-expand. The ratio of  mass flow rates of the primary to the secondary jet ($J=\rho_s A_s u_s/\rho_p A_p u_p$, where $\rho_s=\rho_p$ and $u_s=u_p$) is kept at a constant value of 0.6 by maintaining the air supply at a constant pressure. Firstly, several cases are experimentally demonstrated for a range of low actuation frequencies ($f_a$). Later, unsteady RANS (Reynolds Averaged Navier-Stokes) based computational results are presented to elaborate on the operation extents at a range of high-actuation $f_a$s. A detailed list of cases considered in the present study from experiments and computations are tabulated in Table \ref{tab:exp_comp_cases}. Hereafter, cases under discussion in the subsequent sections are cited through the case number listed in Table \ref{tab:exp_comp_cases}.

\begin{table}
	\caption{\label{tab:exp_comp_cases}Tabulation of dimensionalized and normalized frequencies ($f$ and $fh_j/u_j$) due to the secondary flow's total pressure actuation ($p_{0s}$) at an actuation frequency of $f_a$ from both the experimental and computational cases for flow conditions as mentioned in Table \ref{tab:flow_cond} in the present research of the supersonic fluidic oscillator. Subscript $s$ and $p$ represent the measured frequencies from pressure fluctuations in the exit plane of the secondary or primary jet.}
	\begin{ruledtabular}
		\begin{tabular}{lcc}
			\textbf{Cases} & $f_a=f_s=f_p$ (Hz)\footnote{In the previous work of some of the authors \cite{Maikap2022}, numerically they proved that $f_a=f_s=f_p$, which is also seen in the present experiments and computations with variations less than a percentage.} & $[f_s h_j /u_j] = [f_p h_j /u_j]$\\ \midrule
            \multicolumn{3}{c}{\textbf{Experiments}} \\ \midrule
            C-0\footnote{The primary jet and the secondary jet at the top are turned on completely. There are no transient events in this case as the primary and secondary jet flow are steady ($\dot{m}_s=\dot{m}_{s_t}=\dot{m}_{s_b}$, and $\dot{m}_s \neq f(t)$, where the subscript $s_t$ and $s_b$ indicate the secondary jet flow from the top and bottom chamber).} & - & - \\
            C-I & 0.66 & 1.785$\times 10^{-5}$ \\
            C-II & 1 & 2.7047$\times 10^{-5}$\\
            C-III & 2.5 & 6.7618$\times 10^{-5}$\\
            C-IV & 5 & 13.524$\times 10^{-5}$\\
            C-V & 6.6 & 17.85$\times 10^{-5}$\\ \midrule
            \multicolumn{3}{c}{\textbf{Computations}} \\ \midrule
            C-VI & 16.67 & 0.45$\times 10^{-3}$ \\
            C-VII & 33.33 & 0.9$\times 10^{-3}$\\
            C-VIII & 100 & 2.7$\times 10^{-3}$\\
            C-IX & 150 & 4.1$\times 10^{-3}$\\
            C-X & 250 & 6.8$\times 10^{-3}$ \\
            C-XI & 500 & 13.5$\times 10^{-3}$ \\
            C-XII & 1666.67 & 45.1$\times 10^{-3}$\\
            C-XIII & 5000 & 135.2$\times 10^{-3}$\\
		\end{tabular}
	\end{ruledtabular}
\end{table}

\subsection{Gross flow features} \label{ssec:gross_flow}

\begin{figure*}
	\includegraphics[width=\textwidth]{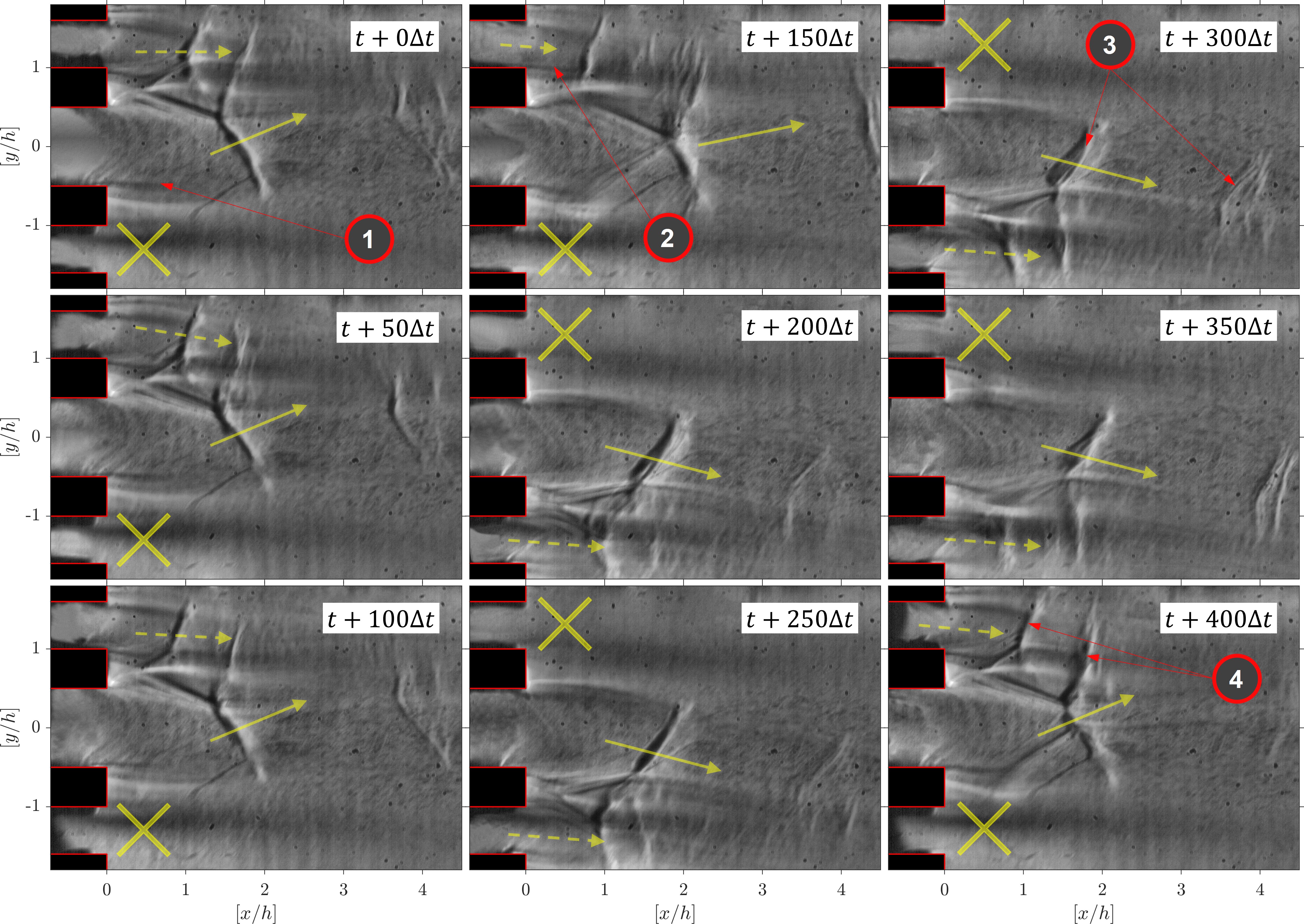}
	\caption{\label{fig:sch_inst}\href{https://youtu.be/gwZ53SN2dsk}{(Multimedia View)} Normalized instantaneous shadowgraph image  snapshots which qualitatively represent the line-of-sight integrated sum of the apparent density double derivatives. The snapshots taken at different time instants ($\Delta t = 1/f_s$, where $f_s=1000$ Hz) show the transverse jet flapping on the $xy$ plane for the case C-II at a jet flapping frequency of $f=1$ Hz. The yellow arrow mark $(\rightarrow)$ represents the tentative deflection and direction of the primary (solid line) and secondary (dash line) jet flow near the $yz$ plane at $[x/h]=0$. The cross mark $(\times)$ indicates the absence of flow in either the secondary flow's top or bottom portion. Flow features: 1-2 under-expanded primary and secondary jet, 3-4 successive distorted Mach stems from the primary and secondary jet.}
\end{figure*}

Figure \ref{fig:sch_inst} shows the shadowgraph images of the sonic jet oscillator flowfield at nine different time instances ($\Delta t = 1/f_s$, where $f_s=500$ Hz). The C-II case with the actuation frequency of $f_a=1$ Hz is considered to understand the gross flow features. The top secondary jet is actuated from $t+ 0\Delta t$ to $t+ 150\Delta t$. As a result, the primary jet is deflected upwards. The primary and secondary jet's directions are shown as arrows with solid and dashed yellow lines, respectively, in Figure \ref{fig:sch_inst}. The under-expanded primary and secondary jet and the respective Mach stems (at least two) from the first and second shock cells are also visible from the imaging between $t+ 0\Delta t$ and $t+ 150\Delta t$. The primary and secondary jet later merge and forms a single jet downstream. The primary jet deflection direction shifts from the top wall side towards the bottom between $t+ 200\Delta t$ and $t+ 350\Delta t$ as the secondary jet actuation ends at the top and a subsequent actuation begins at the bottom.

The moment actuation in the top secondary jet begins again after closing the bottom secondary jet at $t+ 300\Delta t$. Subsequently, the primary jet starts deflecting again towards the top wall. Finally, the primary jet attaches to the top wall, completing an oscillation cycle. It can be seen that this oscillation cycle is repeated for further operation of the fluidic oscillator as shown in the \href{https://youtu.be/gwZ53SN2dsk}{Multimedia View} of Figure \ref{fig:sch_inst}. The shadowgraph experiments of the current fluidic oscillator reveal the following key findings: the streamwise actuation of an under-expanded secondary jet to one side of another under-expanded primary jet in a confined duct can deflect the primary jet to the actuated secondary jet's side comfortably. Hence, the secondary jet can be injected cyclically to oscillate the primary jet in the cross-stream direction.

\subsection{Mach stem oscillations} \label{ssec:shock_osc_stud}

\begin{figure*}
	\includegraphics[width=\textwidth]{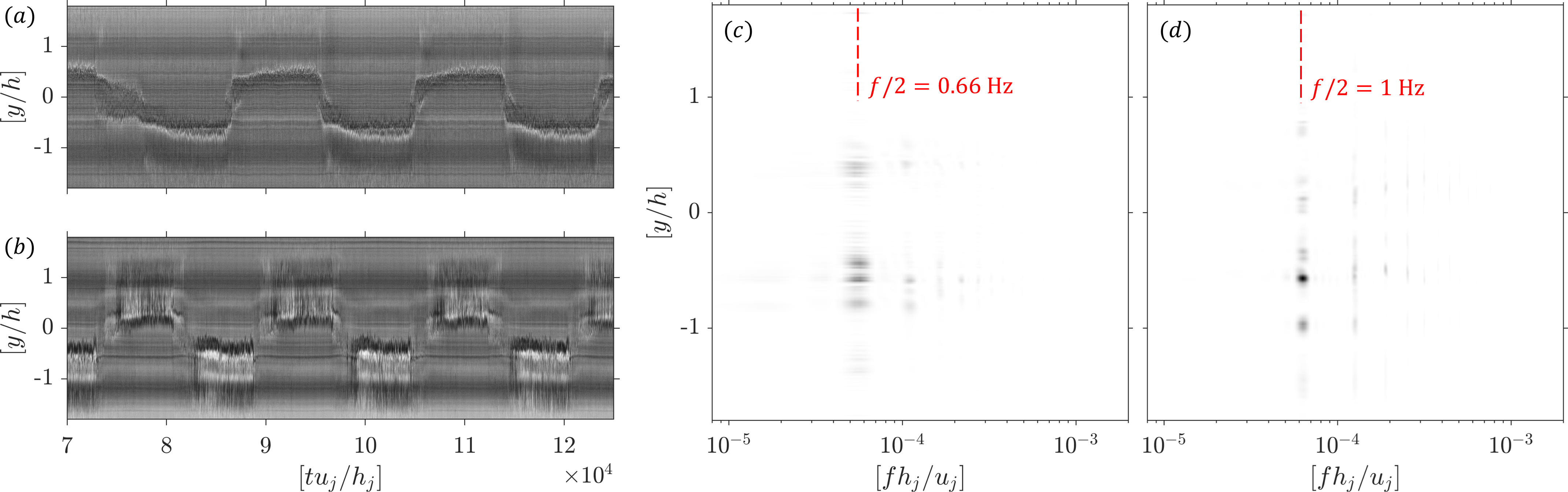}
	\caption{\label{fig:xt_xf_plots}\href{https://youtu.be/rAdb87HgCxY}{(Multimedia View)} Construction of respective $y-t$ and $y-f$ plots from the shadowgraph images for two different cases: C-I and C-II (see Table \ref{tab:exp_comp_cases}). The $y-t$ plot is constructed by stacking the light intensity variation profiles sampled at $[x/h]=1.5$ across different times next to one another. The trace of the transverse jet flapping is seen in the $y-t$ plots (a-b). The corresponding row-wise Fast Fourier Transform (FFT) of the $y-t$ images reveals the dominant spectral contents in the $y-f$ diagram (c-d).}
\end{figure*}

The shadowgraph images shown in the previous section reveal the oscillation of the primary jet towards the test section's top and bottom walls with the controlled jet actuation. The change in flow direction due to jet oscillation also results in oscillation of the Mach stem\cite{Sivaprasad2023} in the transverse direction, as depicted in the shadowgraph images shown in Figure \ref{fig:sch_inst}. As shadowgraph is a line-of-sight light integrated apparent density double derivative, the Mach stems are visualized with a sudden contrast change in the images. Tracking the primary jet's Mach stem displacement spatially inadvertently results in monitoring the primary jet's deflection or response to the secondary jet's actuation. One way to enable tracking the primary jet's Mach stem is to construct a trajectory plot\cite{Karthick2021,Karthick2023} of shadowgraph light intensity\cite{Sugarno2022} along the transverse location at a fixed streamwise location. 

Figure \ref{fig:xt_xf_plots} a-b shows the typical trajectory or $y-t$ plot of Mach stem oscillation in the transverse direction for C-I and C-II cases (see Table \ref{tab:exp_comp_cases}). A fixed streamwise location of $[x/h]=1.5$ is considered, and the pixel intensity variation along the available $[y/h]$ between $-2\leq [y/h] \leq 2$ is collected from the shadowgraph images for each time instant. The collected pixel intensity profiles are stacked along the $x$-axis carrying the flow time stamp ($tu_j/h_j$) to finish constructing the $y-t$ plot. The Mach stem's spatial displacement is distinctly visible as a contrasting line varying in the form of a square or step wave, indicating the primary jet's immediate response to the secondary jet's actuation.

The importance of constructing the $y-t$ plots lies in their ability to perform simplified spectral analysis using Fast Fourier Transform (FFT). The images in Figure \ref{fig:xt_xf_plots} a-b are subjected to FFT analysis to directly yield the $y-f$ plot, which helps in tracking the spectral variations spatially as shown in Figure \ref{fig:xt_xf_plots} c-d. The actuation frequency ($f_a$) for C-I and C-II are 0.66 and 1 Hz, respectively. The primary jet's response spectra ($f_r$) are also found to be the same as identified in the $y-f$ plots shown in \ref{fig:xt_xf_plots} c-d. The final values are computed as half of the original values as the Mach stem crosses a spatial point twice in a single cycle. Moreover, for a discrete $f_a$, $f_r$ tends to carry a multitude of tones, however, at a very low amplitude. The inherent coupling of jet flow and actuation modes might be responsible for producing higher harmonics.

\subsection{Unsteady pressure measurements} \label{ssec:press_meas}

\begin{figure*}
	\includegraphics[width=0.8\textwidth]{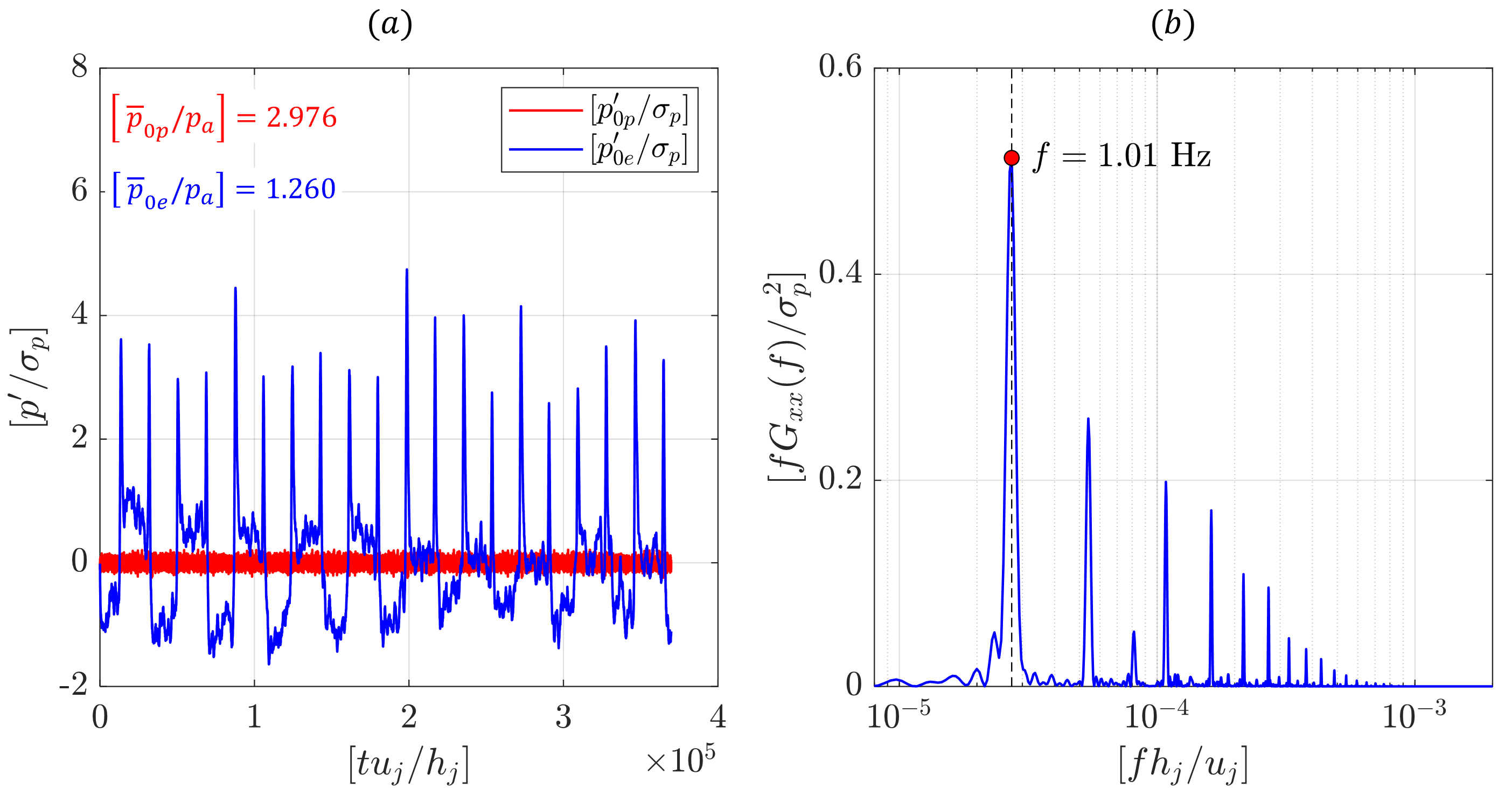}
	\caption{\label{fig:first_readings} (a) Non-dimensionalized Pitot probe pressure fluctuation measurements taken at two different locations: $[x/h,y/h]=[0,0]$ (${p'_{0p}}/p_a$ - solid red line) and $[x/h,y/h]$$=[5,0]$ (${p'_{0e}/p_a}$ - solid blue line) showing the establishment of stationary in-flow conditions and pulsating pressure fluctuations due to transverse jet flapping (see Figure \ref{fig:model_details} for identifying the measurement locations); (b) Non-dimensionalized pre-multiplied power spectral density obtained from $p'_{0e}$ showing the presence of a discrete dominant spectrum ($f=1.01$ Hz) and its recessive harmonics for C-II (see Table \ref{tab:exp_comp_cases}). A standard deviation of $p'_{0e}$ ($\sigma_p = \sigma_{p'_{0e}} = 0.0534$ bar) is used as the reference pressure for normalization.}
\end{figure*}

It should be noted that monitoring the stagnation pressure close to the duct's exit can be used to identify the primary jet oscillation frequency. A carefully placed pitot probe near the duct's exit and the primary jet's centerline ($[x/h,y/h] = [5,0]$) has been therefore used to measure the merged jet's unsteadiness. Figure \ref{fig:model_details}a shows the exit plane marking - \encircle{$e$} where the pitot measurements are done. Figure \ref{fig:first_readings}a shows the non-dimensional pressure fluctuations from the primary jet's settling chamber ($p_{0p}$) and the pitot probe ($p_{0e}$) placed near the test section's exit plane. The pressure fluctuations are derived from the pressure history data obtained from pressure sensors provided at the primary jet settling chamber and the pitot probe near the exit plane. It can be seen from \ref{fig:first_readings}a that the values of $p_{0p}$ remain stationary for the considered flow time carrying minimal broadband fluctuations. However, $p_{0e}$ values exhibit a periodic event with large fluctuations. Figure \ref{fig:first_readings}b shows the spectra of time-varying $p_{0e}$ (the pitot probe pressure fluctuation history from Figure \ref{fig:first_readings}a) computed using the Fast Fourier Transform (FFT) to highlight the existence of a discrete spectrum ($f_p=1.01$ Hz) and its harmonics. It can be seen that the dominant frequency of the pitot probe pressure oscillation spectra matches nearly the secondary control jet actuation frequency. 

\begin{figure*}
	\includegraphics[width=\textwidth]{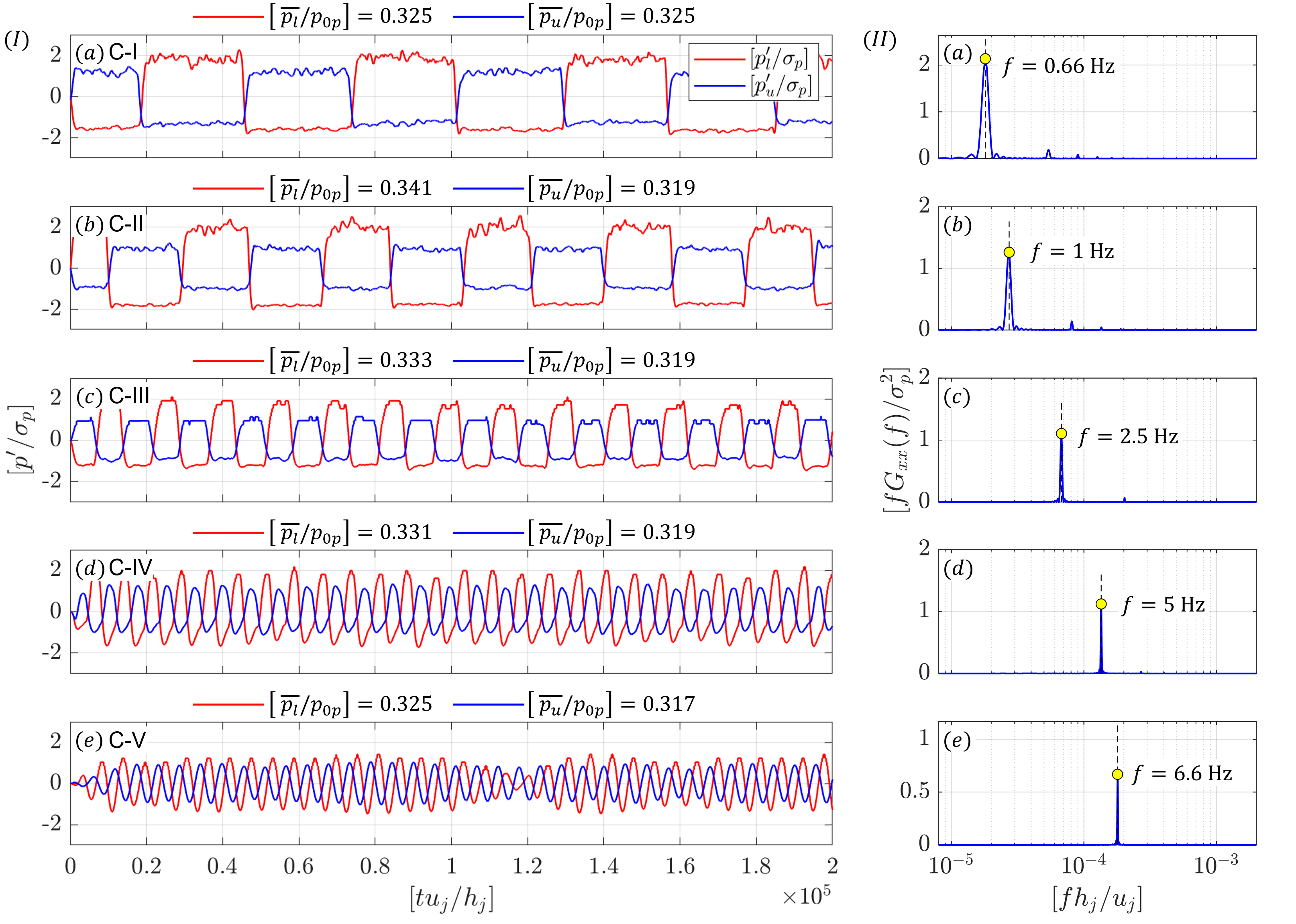}
	\caption{\label{fig:exit_variations}(I) Non-dimensionalized pressure fluctuation measurements taken at two different locations: $[x/h,y/h]=[4,-2.5]$ (lower wall - solid red line) and $[x/h,y/h]$$=[4,2.5]$ (upper wall - solid blue line) showing the pulsating pressure fluctuations due to transverse jet flapping (see Figure \ref{fig:model_details}a for identifying the measurement locations). The mean pressure is mentioned in the legends of the sub-figures; (II) Corresponding non-dimensionalized pre-multiplied power spectral density showing the presence of a discrete dominant spectrum. (a-e) The aforementioned analysis was done for all the experimental cases from C-I to C-V (see Table \ref{tab:exp_comp_cases}). A standard deviation of $p'_{0e}$ ($\sigma_p = \sigma_{p'_{0e}} = 0.0534$ bar) is used as the reference pressure for normalization.}
\end{figure*}

Figure \ref{fig:exit_variations}-I show the normalized unsteady wall-static pressure fluctuations at stations \encircle{$u$} (solid blue line) and \encircle{$l$} (solid red line) as shown in the schematic of Figure \ref{fig:model_details}a for the experimental cases ranging from C-I to C-V (see Table \ref{tab:exp_comp_cases}). Almost 7 to 35 oscillation cycles are captured across the experimental cases during the considered stationary flow time. Each oscillation cycle resembles a square wave or a step profile which is seen in the cases with low $f_a$s as in C-I (Figure \ref{fig:exit_variations} I-a). A rapid rise in pressure fluctuations ($p'$) corresponds to the merged jet's sudden attachment to the respective wall side where the measurements are taken. Similarly, a rapid dip in $p'$ indicates the merged jet's detachment from the wall. The observation of constant $p'$ in each step cycle corresponds to the time when the merged jet remains attached to the respective wall. The frequent changes in the $p'$ trend indicate that the jet deflection is rapid and quickly responds to the secondary jet's actuation. Figure \ref{fig:exit_variations}-II carry the spectra of the $p'$ time series measured at station \encircle{$u$}. As the secondary jet's actuation frequency ($f_a=f_s$) increases, the merged jet's response ($f_r=f_p$) also increases and matches that of the corresponding $f_a$. In the previous work carried out by the authors\cite{Maikap2022}, the above observation is verified numerically using unsteady RANS.    

\subsection{Sonic jet oscillator's flow physics} \label{ssec:flow_phys}

When the secondary jet actuates on one side, the primary jet is pulled towards the secondary jet's actuation side. Later, both primary and secondary jet merges and deflects as one. The deflecting merged jet  entrains the available fluid mass near the respective wall. The rapid entrainment creates low pressure in the proximity causing the merged jet to stick to the wall. Such a phenomenon is called the `Coanda effect'. Once the secondary jet's actuation switches sides, the merged jet also turns to the other side wall and sticks instantly. As the secondary jet's actuation frequency varies, the primary jet responds linearly to the variation. The events are previously qualitatively shown through the shadowgraph experiments in Sec. \ref{ssec:gross_flow}. The linear response between the actuation and response is further verified by constructing $y-t$ and $y-f$ plots.  

\subsection{Operational capabilities at high actuation frequencies} \label{ssec:num_study}

\begin{figure*}
	\includegraphics[width=\textwidth]{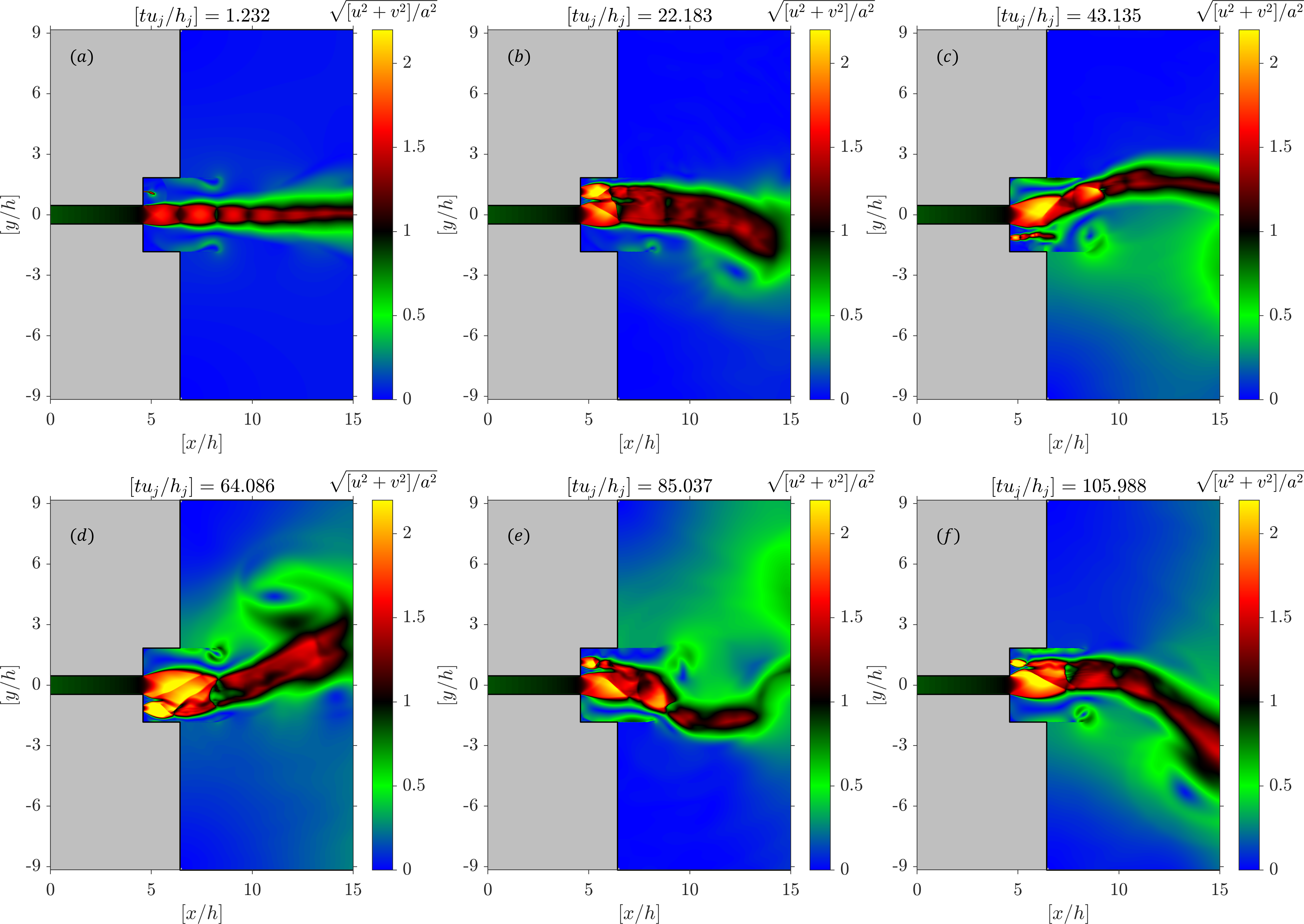}
	\caption{\label{fig:cfd_contours}\href{https://youtu.be/6lv4zNN2Xr8}{(Multimedia View)} (a-f) Normalized instantaneous contours of velocity magnitude reveal the transverse jet flapping in the supersonic fluidic oscillator at different normalized time instants for the C-XI case (see Table \ref{tab:exp_comp_cases}). The flow is from the left to the right. The velocity magnitude is normalized by the local sound speed at that temperature. The secondary jet in the top portion is turned on in $b,e,$ and $f$, whereas the bottom portion of the secondary jet is turned on in $c$ and $d$. The black colour contour lines represent the sonic Mach number or $\left[(u^2+v^2)/a^2\right]^{0.5}=1$.}
\end{figure*}

 A large flow rate facility is utilized to study the present problem with excellent spatial resolution. Moreover, the resulting adequate test section size enables quality flow visualization and provision to mount sensors. One setback from the current design's relatively large test section size is the solenoid valve's utilization. It has a significant response time to actuate ($t_r >150$ ms) owing to the high mass flow rate it handles. The limiting speed in opening the solenoid valve primarily affects the operational range of the secondary jet's actuation frequency ($f_a$). Hence, $f_a$ for the present experimental study is limited to only $0.66 \leq f_a \leq 6.6$ Hz (see Table \ref{tab:exp_comp_cases}). A broad computational campaign is sought to study the supersonic jet oscillator's operational characteristics at frequencies higher than what is being discussed in the experimental section. The oscillating jet flow is spatiotemporally resolved by solving the unsteady RANS equations as detailed in Sec. \ref{sec:num_meth}. Some of the vital results are discussed in the present section.

\begin{figure*}
	\includegraphics[width=0.8\textwidth]{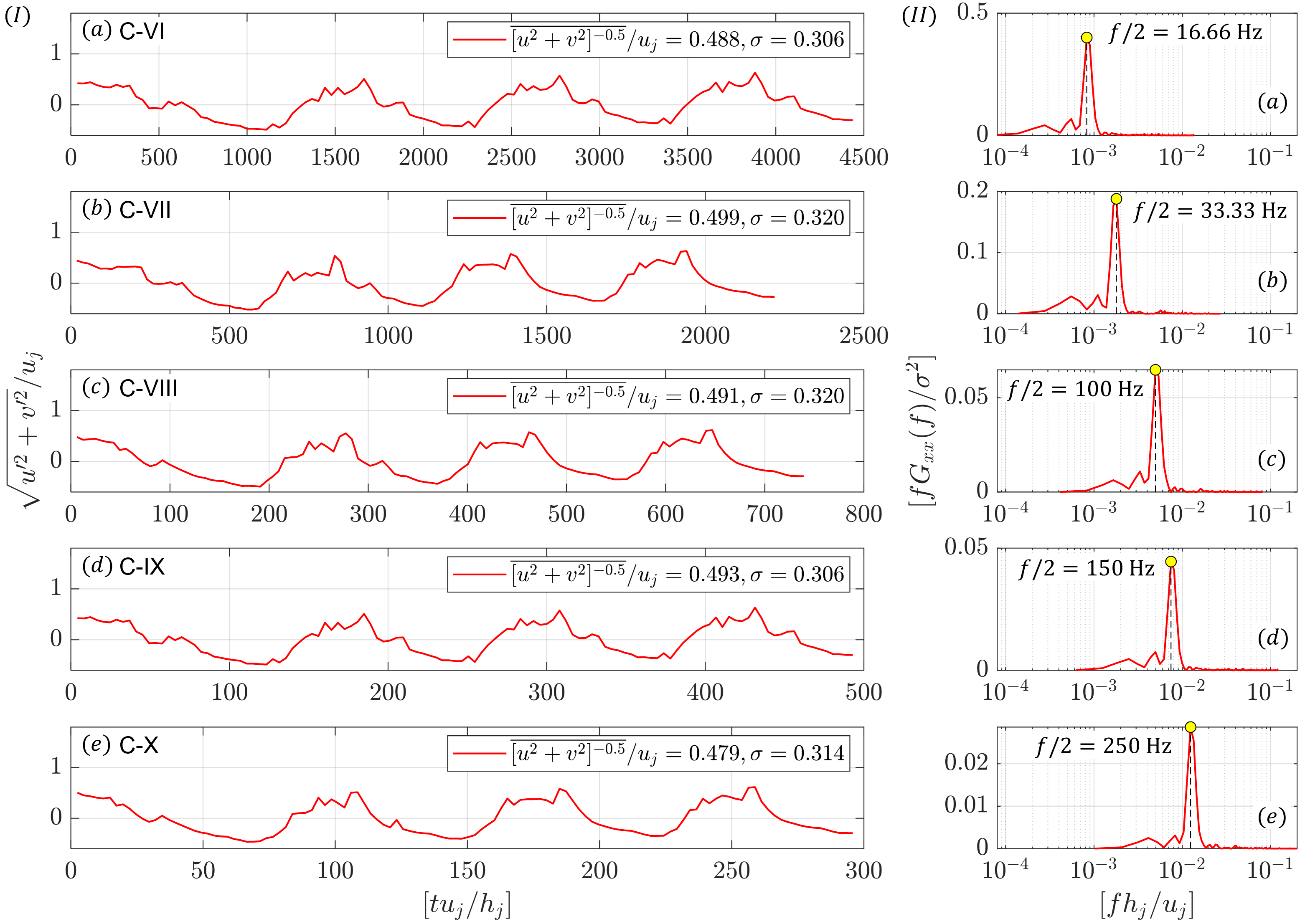}
	\caption{\label{fig:cfd_spectra}(I) Non-dimensionalized velocity magnitude fluctuation measurements taken at $[x/h,y/h]=[5,0]$ (solid red line) showing the pulsating velocity field fluctuations due to transverse jet flapping (see Figure \ref{fig:model_details} for identifying the measurement locations). The mean and standard deviation of the velocity field is given in the legends of the sub-figures; (II) Corresponding non-dimensionalized pre-multiplied power spectral density showing the presence of a discrete dominant spectrum. (a-e) The aforementioned analysis was done for all the computational cases from C-VI to C-XI (see Table \ref{tab:exp_comp_cases}). As the jet flaps about the flow axis twice, the dominant spectrum is half of its original value.}
\end{figure*}

Figure \ref{fig:cfd_contours} shows the velocity magnitude contours at different time instants for the C-XI case where the secondary jet has $f_a = 500$ Hz, an order of magnitude higher than the experiments. At $[tu_j/h_j]=1.232$, the secondary jet actuation is absent, and the primary jet is exhausted to the ambient without any perturbations. At $[tu_j/h_j]=22.183$, the top secondary jet is actuated, leading to the deflection of the primary jet to the upper wall. At $[tu_j/h_j]=43.135$ and $[tu_j/h_j]=64.086$, the secondary jet at the bottom side is actuated, resulting in the primary jet's adherence to the lower wall. From $[tu_j/h_j]=85.037$, the cycle repeats, and the primary jet is shown to be oscillating continuously. The readers are referred to the \href{https://youtu.be/6lv4zNN2Xr8}{Multimedia View} given in Figure \ref{fig:cfd_contours} to appreciate the primary jet's sustained cycles at high $f_a$.

The quantitative nature of the merged jet oscillation characteristics is extracted by measuring the velocity magnitude at $[x/h,y/h]=[5,0]$ for all the cases whose $f_a$ vary between $16.67 \leq f_a \leq 250$ Hz. Three consecutive oscillation cycles are plotted for the cases ranging from C-VI to C-X in Figure \ref{fig:cfd_spectra}-I. The signals are later subjected to FFT analysis as shown in Figure \ref{fig:cfd_spectra}-II. The spectral amplitude is normalized using the local signal's square of the standard deviation. It should be noted that the frequency obtained from the FFT is twice the jet oscillation frequency. As the jet passes the midpoint twice every cycle, the velocity magnitude shows the maximum value (peak) two times. Hence, the frequency from velocity fluctuation is identified as double the oscillation frequency. Moreover, the spectra also verify the merged jet's response is coherent with the secondary jet's actuation. Therefore, the present control strategy is shown to be efficient for a wide range of frequencies, between $0.66 \leq f_a \leq 500$. 

\subsection{Limiting operating conditions} \label{ssec:lim_op}

\begin{figure*}
	\includegraphics[width=0.8\textwidth]{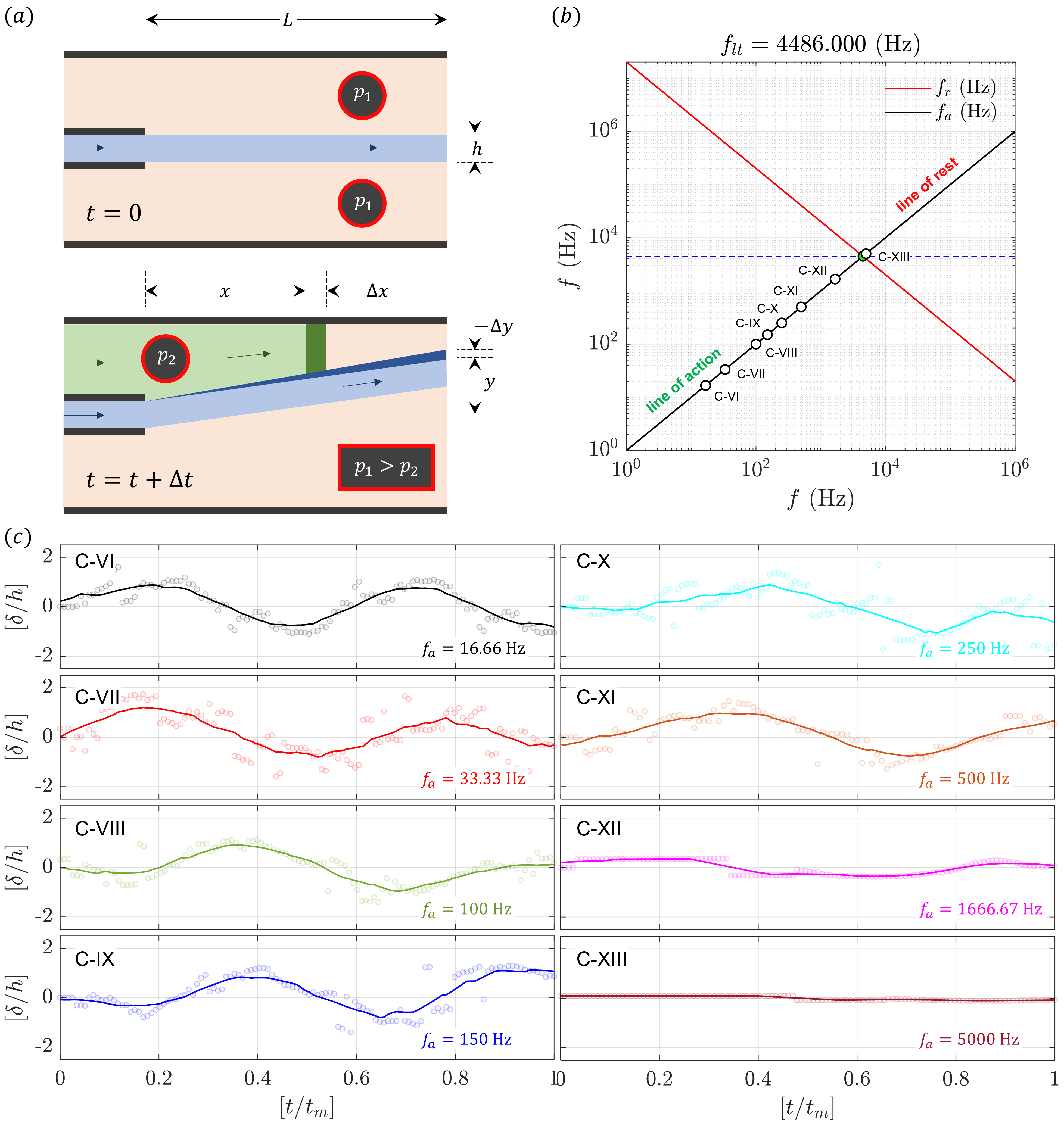}
	\caption{\label{fig:limit_schem} (a) Schematic representation of the  proposed reduced-order model for the supersonic jet oscillator: the top figure shows the initial state ($t=0$) of the fluidic block when both the top and bottom actuation are absent; the bottom figure represents the state of the fluidic mass at time $t=t+\Delta{t}$ when the secondary jet at the top is actuated. (b) Graphical plot depicting the limiting operating conditions for the considered supersonic jet oscillator whose parameters are listed in Table \ref{tab:flow_cond}, Figure \ref{fig:model_details}, and Figure \ref{fig:exit_variations}. The black solid line indicates the secondary jet's possible actuation frequencies ($f_a$), whereas the red solid line shows the primary jet's response frequencies ($f_r$) for the range of frequencies considered. The circular markers indicate the present cases considered in both experiments and computations; (c) Line plots showing the transverse jet deflection height variation ($\delta/h$) measured at the exit plane at different normalized time steps ($t/t_m$, where $t_m$ is the maximum time) from the numerical simulations of all the cases between C-VI to C-XIII. The markers indicate the computed $[\delta/h]$ from a specific condition, and the solid line represents the global smoothed trend.}
\end{figure*}

It is seen from the previous sections that the response frequency ($f_r$) of the primary jet is dependent on the secondary jet's actuation frequency ($f_a$). However, when $f_a$ is increased to a substantial value, the performance of the current fluidic oscillator will start to decline and eventually cease. The jet fluid mass limits the response of the oscillator to the available pressure difference generated by the secondary jet. Hence, a tentative estimate of the limiting frequency ($f_{lt}$) is needed to provide the device's operational extrema. One way to arrive at $f_{lt}$ is by reduced-order modelling of the flowfield, where a mass of the fluid block (primary jet) oscillates due to the secondary jet's actuation on the top and bottom of the fluid block. Such a reduced-order analysis may not model the entire flow physics. However, deploying well-known engineering mechanics relations could easily estimate the limiting jet response condition to the first order. Figure \ref{fig:limit_schem}a represents a schematic where the supersonic oscillator is simplified using the aforementioned reduced-order model. The pale orange colour indicates the sub-atmospheric pressure ($p_1$), and the pale blue and green colours indicate primary and secondary jet flow. 

In Figure \ref{fig:limit_schem}a, the primary jet is modelled as a one-sided-hinged mass of rigid block about the nozzle exit. The fluid block carries a uniform mass of $m$ spanning the duct length of $L$ and height $h$. The duct has a rectangular cross-section of height $H$ and width $w$. The duct's top and bottom portions allow the secondary jets to expand and propagate in the streamwise direction. The propagation creates a pressure drop less than the sub-atmospheric condition ($p_2 < p_1$) due to the Coanda effect in the region where the secondary jet is actuated. Meanwhile, a higher pressure region ($p_1>p_2$) is present on the other side, where the secondary jet is absent, thereby maintaining a net differential pressure of $\Delta p=p_1-p_2$ across the fluidic block. The established pressure differential causes the block of fluid mass to deflect towards the side where the secondary jet is actuated. The top-side deflection of the primary jet is considered to formulate the governing differential equations. Newton's second law of motion is used to identify $f_{lt}$. The force acting on the fluidic block in the transverse direction ($y$) is given as, 
\begin{equation}
    F_y = m \times a_y, \label{eq:m0}
\end{equation}
Where $F_y$ is the transverse force acting on the fluidic block, and $m$ and $a_y$ are the block's mass and acceleration along the $y$-direction, respectively. 

Figure \ref{fig:limit_schem}a shows a schematic representation of the fluid block's dynamic state. At $t=0$, the primary jet remains undeflected, and the secondary jet is absent. At $t=t+\Delta t$, the top side's secondary jet is actuated. As the secondary jet propagates in the streamwise direction ($x=x+\Delta x$), the pressure gradient across the fluidic block increases leading to the eventual primary jet's displacement ($y=y+\Delta y$) to the top side. The left-hand side of Eq. \ref{eq:m0} is further represented in terms of the differential pressure ($\Delta p$) acting on the primary jet block's planar area of $w \times x$ at time $t$ as shown in Eq. \ref{eq:m1}. 
\begin{equation}
    F = \Delta p \times w \times x. \label{eq:m1}
\end{equation}
Here, the weight of the fluidic block (gravity) is considered negligible.

Furthermore, the primary block's mass is written in terms of density ($\rho^*$) and volume ($L \times h \times w $). Here $\rho^*$ and $h$ are the density at the jet exit and height of a non-decaying jet spanning the duct length $L$. Therefore, the equation of force balance on the primary jet block at time instance $t$ can be written as,
\begin{equation}
    \Delta p \times w \times x = \rho^* \times L \times w \times h \times \frac{d^2y}{dt^2} \Bigg\rvert_{t}. \label{eq:m2}
\end{equation}

After an infinitesimally short time $\Delta t$, the secondary jet block moves $\Delta x$ streamwise, and the primary jet block moves $\Delta y$ in the transverse direction. The corresponding force balance equation for the primary jet block at time instant $t= t+\Delta t$ can be written as,
\begin{equation}
   \Delta p \times w \times (x+\Delta x) = \rho^* \times L \times w \times h \times \frac{d^2y}{dt^2} \Bigg\rvert_{t+\Delta t}. \label{eq:m3}
\end{equation}

Subtracting Eqs. \ref{eq:m2} and \ref{eq:m1}, and dividing the difference by $\Delta t$ gives,
\begin{equation}
   \Delta p \times w \times \frac{dx}{dt} = \rho^* \times L \times h \times \frac{d^3y}{dt^3} \Bigg\rvert_{t+\Delta t}. \label{eq:m4}
\end{equation}
Here, $\Delta x/\Delta t \rightarrow dx/dt$ as $\Delta x$ and $\Delta t$ are infinitesimally small values.

Since the secondary jet's opening and closing are carried out in steps (15 distinct steps), the secondary jet's streamwise propagation is assumed as a linear function of $x$. Therefore, the secondary jet's propagation speed is computed as ${dx}/{dt}= {4 \times L}/{t_a}$, where ${t_a}$ is the secondary jet's actuation time (top). Substituting the value of the secondary jet's propagation speed in Eq. \ref{eq:m4} gives,
\begin{equation}
   \Delta p \times w \times \frac{2 \times L}{t_a} = \rho^* \times L \times w \times h \times \frac{d^3y}{dt^3} \Bigg\rvert_t. \label{eq:m5}
\end{equation}
It should be noted that $t_a$ is a constant for any particular case. Therefore, simplification of Eq.\ref{eq:m5} results,
\begin{equation}
    \frac{d^3y}{dt^3} \Bigg \rvert_t = \frac{4 \times \Delta p}{\rho^* \times h \times t_a}.\label{eq:m6}
\end{equation}

Upon integrating Eq. \ref{eq:m6} thrice with respect to $t$ yields,
\begin{equation}
    y= \left[\frac{4 \times \Delta p}{\rho^* \times h \times t_a}\right] \left(\frac{t^3}{6}\right) + C_1\left(\frac{t^2}{2}\right) + C_2(t) + C_3, \label{eq:m9}
\end{equation}
where $C_1$, $C_2$, and $C_3$ are integration constants. It is also known that the initial position of the fluid block is at the duct's centerline during $t = 0$, having zero velocity and acceleration. Thus, at $t = 0$, $y = 0$,  ${dy}/{dt} = 0$, and  ${d^2y}/{dt^2} = 0$. Substituting these boundary conditions, the values of the constants are evaluated as $C_1=C_2=C_3=0$. Therefore, Eq.\ref{eq:m9} can be rewritten as,
\begin{equation}
    y= \left[{\frac{4 \times \Delta p}{\rho^* \times h \times t_a}}\right] \left(\frac{t^3}{6}\right).\label{eq:m10}
\end{equation}

Eq.\ref{eq:m10} expresses the transverse deflection of the fluidic block at time $t$. Further, it can be safely stated that the jet has responded to the actuation when it reaches the maximum possible $y$ location ( $y_{max} = {H-h}/{2}$ ) with a response time of $t_r$. Substituting $y_{max}$ and $t_r$ in Eq.\ref{eq:m10}, the expression for $t_r$ is obtained as,
\begin{equation}
    t_r= \sqrt[3]{ \frac{4 \times (H-h) \times \rho^* \times h \times t_a }{3 \times \Delta p}}. \label{eq:m11}
\end{equation}

In terms of response frequency ($f_r$), the expression in Eq.\ref{eq:m11} is rewritten as,
\begin{equation}
    f_r= \sqrt[3]{\frac{ 2 \times \Delta p \times f_a }{3 \times (H-h) \times \rho^* \times h }}. \label{eq:m12}
\end{equation}

If $f_a < f_r$, the primary jet responds to the secondary jet's actuation. However, the extent of jet deflection progressively vanishes as $f_a \rightarrow f_r$. The jet becomes completely unresponsive once $f_a > f_{lt}$. Thus, the limiting frequency exists when $f_r = f_a$. 
Putting $f_r=f_a$ in Eq. \ref{eq:m12}, the expression for limiting frequency $f_{lt}$ is obtained as,
\begin{equation}
    f_{lt}= \sqrt{ \frac{2 \times \Delta p }{3 \times (H-h) \times \rho^* \times h}}. \label{eq:m13}
\end{equation}

The terms on the right-hand side of Eq. \ref{eq:m13} are evaluated from the present experimental geometry and jet operating conditions. The considered values are listed as follows: $\Delta_p = p_a - p_e = 1\times 10^4$ Pa (from Figure\ref{fig:exit_variations}), $\rho^* = 2.2088$ kg/m$^3$ (throat condition in a sonic jet from isentropic relation and Table \ref{tab:flow_cond}), $h=0.01$ m, and $H=0.04$ m. Thus, the limiting actuation time ($t_{lt}$) is estimated as $0.31527 ms$, and the limiting frequency ($f_{lt}$) is found to be 3.172 kHz. The values are also reflected well in the plot shown in Figure\ref{fig:limit_schem}b. Here, the red line represents the variation of $f_r$ with respect to $f_a$ obtained from Eq.\ref{eq:m12}, and the black line represents the linear variation of $f_a$. The two lines intersect at the location where $f_r =f_a = f_{lt}=3.172 $ kHz. The zone of $f_a$ from the left to the intersection point at $f_{lt}$ is called the `line-of-action', and the oscillator is expected to perform. Any $f_a$ to the right of the intersection point at $f_{lt}$ is called the `line-of-rest' where the primary jet will not deflect to the secondary jet's actuation and remains unresponsive.

The validity of the present model is further tested by running two extra numerical cases (C-XII and C-XIII, see Table \ref{tab:exp_comp_cases}) at higher $f_a$. Later, the jet deflection extent ($\delta/h$) is computed at the duct exit ($x/L=5$) by monitoring the peak velocity magnitude along the $y$ direction for all the numerical cases mentioned in Table \ref{tab:exp_comp_cases} and plotted in Figure \ref{fig:limit_schem}c. As $f_a$ increases, $[\delta/h]$ is pronounced firm. However, as $f_a$ approaches $f_{lt}$, variations in $[\delta/h]$ are small as predicted by the reduced-order model given in Figure \ref{fig:limit_schem}b.

\section{Conclusions}\label{sec:conclusions}
The present study investigates an actively controlled supersonic fluidic oscillator using experimental and numerical methods. The oscillator geometry constitutes a primary jet expanding into a duct with larger dimensions. Two secondary jets emerge from the backwards-facing steps formed due to the sudden enlargement of the duct area and expands parallel to the primary jet. The secondary jets are alternately pulsed in the streamwise direction. It is found that with the actuation of the secondary jet, the primary jet merges with the secondary jet and sticks to the closest wall in the direction of the secondary jet. Alternative actuation of the top and bottom secondary jets creates an oscillating primary jet. A parametric study of the primary jet oscillation frequency with respect to the secondary jet actuation frequency concluded that the primary jet oscillates with nearly the same frequency as the secondary jet actuation cycle frequency. Unsteady shock oscillation characteristics have been determined by studying the angular oscillation of the Mach stem with respect to the centerline. It is seen that the Mach stem oscillation frequency matches the secondary jet actuation frequency. Numerical studies with higher secondary jet actuation frequencies have been carried out to study the effectiveness of the fluidic oscillator at higher frequency ranges. The study showed that the current fluidic oscillator provides a wide range of oscillation frequencies (0.66 Hz to 500 Hz) for supersonic jets. An analytical model was established to predict the limiting frequency of the fluidic oscillator used in the present study and further verified using computational simulations.  
\section*{Acknowledgment}
The authors gratefully acknowledge the financial support for this project from the Aeronautics Research Development Board (ARDB), Defense Research and Development Organization (DRDO) India under Grant number-1960. The second author thanks the Scientific Engineering and Research Board - Department of Science and Technology (SERB-DST), India, for the generous financial assistance in the form of Ramanujan Faculty Fellowship (Dept. Mech. Engg., IITM, Chennai) under grant number RJF/2022/000084.
 
\section*{Authors' Declaration}
The authors declare no conflict of interest.

\section*{Data availability statement}
The data supporting this study's findings are available from the corresponding author upon reasonable request.
\section*{References}
\bibliography{References}
\
\onecolumngrid
\PRLsep
\end{document}